\documentstyle{article}
\begin{document}

\title{The egg-carton Universe}

\author{E. Battaner and E. Florido} 
\maketitle

\begin{bf}
The distribution of superclusters in the Local Supercluster
neighbourhood presents such a remarkable periodicity that some kind of
network must fit the observed large scale structure. A three dimension
chessboard has been suggested$^{1}$. The existence of this
network is really a challenge for currently-suggested theoretical
models. 
For instance, CDM models of the formation of the large scale structure
predict a random distribution of superclusters$^{2}$. If the
filaments of matter that are now observed building up the network are
fossil relics of over-dense regions of magnetic field energy before
Recombination, then it has been shown$^{3}$ that the simplest network
compatible with magnetic field constraints is made up of octahedra
contacting at their vertexes. This suggests a set of superimposed
egg-carton structures. Our aim in this paper is to show that the real
large-scale structure is actually fitted by the theoretical octahedron
structure.
\end{bf}

\vskip 1cm

For this task, a systematic statistical procedure is ``a
priori'' to be preferred. However, such a procedure is difficult to develop,
and the natural human ability to recognize structures provides a
faster search, even if it introduces a degree of subjectivity. In
this case, however, the identification of real octahedra is so clear
and the network is so noticeably well defined that a direct inspection
is straightforward.

A fundamental plane of the egg-carton network would contain a large
number of filaments and therefore a large number of superclusters. One
of these fundamental planes can be identified with the SGZ=0 plane. In
a plane very close to this one the high periodicity in the
distribution of matter was discovered$^{2}$ and a high density of
superclusters is to be found$^{4}$. This means in practice that the plane
of the Local Supercluster coincides with this fundamental plane. 
Fig. 1 shows the $10^{-3}$ clusters Mpc$^{-3}$ contour in the
plane SGZ=0 from Tully et al$^{4}$. The identification of a
fundamental direction within this plane is straightforward. There is a
noticeable alignment passing through Draco, Ursa Major, Leo, Hercules
and the Great Attractor, and a long chain of smaller clusters ending
at Tucana. Another fundamental direction perpendicular to this is also
easily identified in the line connecting the elongated Shapley
Concentration, Hercules, the Great Attractor and Perseus-Pegasus. Other
details in this map enable the obtention of the octahedron side, $a$,
of about 150 $h^{-1}$ Mpc ($h = H_0/100$). This is higher than
the period of 130 $h^{-1}$ Mpc found by Broadhurst et al$^{1}$
(hereafter BEKS), but  the BEKS probe line
cuts our structure at length intervals shorter than the octahedron
side. At planes $SGZ = a/\sqrt{2}$ (half a diagonal of the octahedra,
about 106 $h^{-1}$ Mpc) and $SGZ = -a/\sqrt{2}$ the
other vertexes of the identified octahedra are to be found. 
Other planes at $SGZ =
na\sqrt{2}$ (with $n$ being an integer) would contain other fundamental
planes parallel to $SGZ =0$.

Figure 2 provides a schematic view of the identified
octahedra. $A$ and $B$ belong to a region for which supercluster
catalogues are complete. $C$ and $D$ are also well identified and some
vestiges of $A'$ and $B'$ can also be appreciated. Vertexes approximately
contained in the $SGZ=0$ plane are called $A1$, $A2 \dots$, $B1$, $B2
\dots$. Vertexes approximately contained in the $SGZ =a/\sqrt{2}$
plane are called $A5$, $B5$, $C5$, $D5$ and vertexes approximately
contained in the $SGZ =-a\sqrt{2}$ plane are called $A6$, $B6$, $C6$,
$D6$.

Virtually the whole sample of superclusters and voids does, in fact, match the
theoretical egg-carton structure. To identify the network structure,
we used the supercluster catalog from Einasto et al$^{5}$ (hereafter
ETJEA) and the void catalog by Einasto et al$^{6}$ (hereafter
EETDA).

\vskip 1cm

\noindent {\bf{Identification of superclusters}}

A1 $\equiv$ extension of the Virgo-Coma supercluster. A2 $\equiv$
ETJEA 127. A3 $\equiv$ Hydra-Centaurus. A4 $\equiv$ Ursa Maior. A5
$\equiv$ ETJEA 154. A6 $\equiv$ Sextans. Edge A2A3 $\equiv$ Shapley
concentration; Edge A3A4 $\equiv$ Leo; Edge A1A2 $\equiv$ ETJEA 126;
Edge A1A4 $\equiv$ Virgo-Coma; Edge A3A5 $\equiv$ Hercules.

B1 $\equiv$ A3; B2 $\equiv$ ETJEA 16 + Grus-Indus; B4 $\equiv$ Pisces;
B5 $\equiv$ Aquarius-Cetus; B6 $\equiv$ Horologium-Reticulum. Edge
B3B4 $\equiv$ Piscis-Cetus. Edge B1B6 $\equiv$ Phoenix. Edge B4B5
$\equiv$ Perseus-Pegasus. C1 $\equiv$ B3. C5 $\equiv$ ETJEA 207. C6
$\equiv$ Fornax. Edge C1C2 $\equiv$ Sculptor + ETJEAS. D2 $\equiv$
Tucana. D4 $\equiv$ C2.

There are other superclusters matching the net not contained in the
plotted octahedra A, B, C, D. Draco lies in the next vertex in the
direction A3A4. Leo A is at the lower point in the octahedron before
A. Over A1, ETJEA 154 is found at the next point. Piscis-Aries is at the
edge extrapolating B1B4. ETJEA 63 lies below B2. Fornax-Eridanus is
found below B3 in the next
octahedron. Above B3, there is Aquarius B. Edge
B2D1 $\equiv$ ETJEA 6. Microscopium is at the edge above
B2D1. Aquarius-Capricornio is above B3B5. Aquarius B, is above B3B5. All
these perfectly match the proposed net.

All important superclusters are included in the above list, with the
possible exceptions of Leo A, Bootes and Grus. We interpret the above
as meaning that Aquarius
would correspond to the vertex above B3, but that here the net has becomed
deformed due to the huge gravitational attraction produced by the
Piscis-Cetus large mass. The fundamental plane is also gravitationally
deformed by Piscis-Cetus. Under this interpretation, the larger
concentration found in the SGY =0 plane$^{6}$ would be associated with
the large Piscis-Cetus attraction.

\vskip 1cm

\noindent {\bf{Identification of voids}}

In accordance with the above description, there are two kinds of voids:
intra-octahedric and inter-octahedric voids. Connection must exist
between all of them, as a network of filaments is being considered, but
especially between inter-octahedric voids. The following numbering corresponds to
the number in the EETDA void catalog.
1 $\equiv$ inside B. 2 $\equiv$ below B3B4. 3 $\equiv$ below B2. 4
$\equiv$ below B3. 5 $\equiv$ inside B. 6 $\equiv$ below B3B4. 7
$\equiv$ below B4 (though too low; this is the same deformation
induced by Piscis-Cetus). 8 $\equiv$ inside the octahedron below B. 9
$\equiv$ below A3, in the South Local Void, SLV. 10 $\equiv$ inside
A'. 11 $\equiv$ on the line A6A'6. 12 $\equiv$ below edge A1A4. 13
$\equiv$ inside the octahedron below A. 14 $\equiv$ inside A, somewhat
too
low. 15 $\equiv$ below A2. 16 $\equiv$ below A1A4. 17 $\equiv$ below
A1A2. 18 $\equiv$ inside A. 19 $\equiv$ inside A. 20 $\equiv$ above
A1A4, is Bootes Void. 21 $\equiv$ above A3A4. 22 $\equiv$ above
A'1A'2. 23 $\equiv$ above A1A4. 24 $\equiv$ above A3, is the North
Local Void, NLV. Only 25, 26 and 27 do not perfectly match the
structure.

Therefore, though very massive concentrations like that of
Piscis-Cetus may deform the net, it is very clearly identifiable and
previously studies that detected regularities and periodicities are in
agreement and explained by the 3D picture of this egg-carton
network. Magnetic field inhomogeneities with  typical lengths greater
than the horizon along the radiation dominated era are able to
explain this network. Therefore, very large-scale magnetic fields may
have played a very important role in building up the present large-scale
structure of the Universe.

\vskip 1cm
Knowledgements. This paper has been supported by the spanish ``Ministerio de Educacion
y Cultura'' (PB96-1428) and the ``Plan Andaluz de Investigacion''
(MFQ-0108).

\vskip 1cm
\noindent References

(1) Broadhurst, T.J., Ellis, R.S., Koo, D.C. and Szalay, A.S. (1990)
Nature 343, 726-728

(2) Einasto, J., Einasto, M., Gottl$\ddot{o}$ber, S., M$\ddot{u}$ller,
V., Saar, V., Starobinsky, A.A., Tago, E., Tucker, D., Andernach, H.,
Frisch, P. (1997) Nature 385, 139-141

(3) Battaner, E., Florido, E., Jimenez-Vicente, J. (1997) A\&A 326,
13-22; Florido, E., Battaner, E. (1997) A\&A 327, 1-7;
Battaner, E., Florido, E., Garcia-Ruiz, J.M. 1997, A\&A, 327, 8-10

(4) Tully, R.B., Scaramella, R., Vettolani, G., Zamorani, G. (1992)
ApJ 388, 9-16

(5) Einasto, M., Tago, E., Jaaniste, J., Einasto, J. \& Andernach,
H. 1997, A\&AS, 123, 129-146

(6) Einasto, M., Einasto, J., Tago, E., Dalton, G.B., Andernach,
H. (1994) MNRAS 269, 301-322

\newpage

Legends for figures:

Figure 1.- The octahedron network in a fundamental plane nearly
coincident with the SGZ = 0 plane superimposed to the $10^{-3}$
clusters MPC$^{-2}$ contour from Tully et al.$^4$. Units for SGX and
SGY should be multiplied by $h^{-1}$. The obscuration zone and the
Broadhurst et al.$^1$ probe line are also shown.

Figure 2.- Schomatic plot of identified octahedra. A, B, C and D are
the observed octahedra. Points 1, 2, 3 and 4 are in the SGZ = 0
plane. Points 5 lie over the sheet plane. Points 6 lie under the sheet
plane. The axes in this figure would be similar to those in fiure 1.

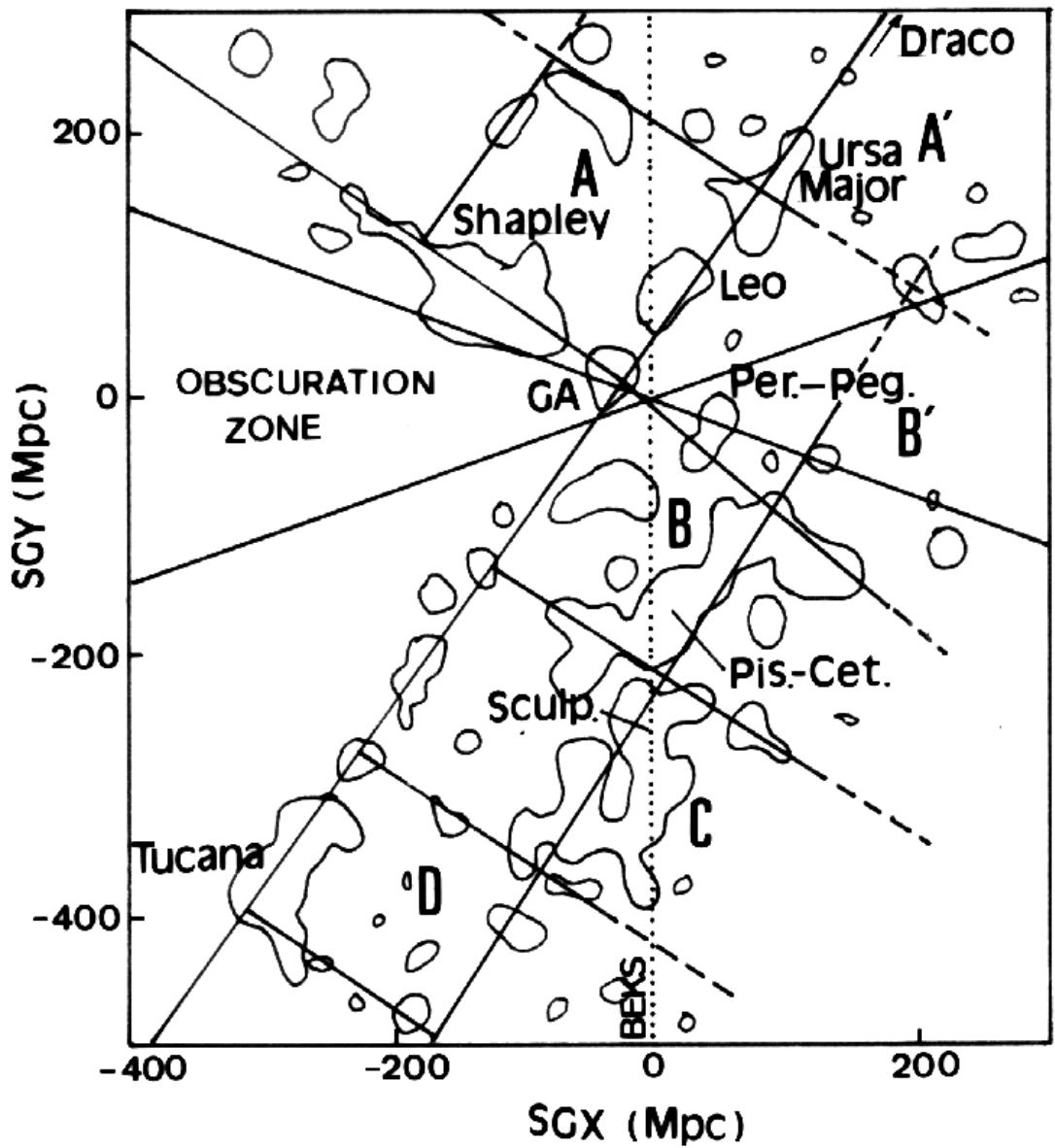
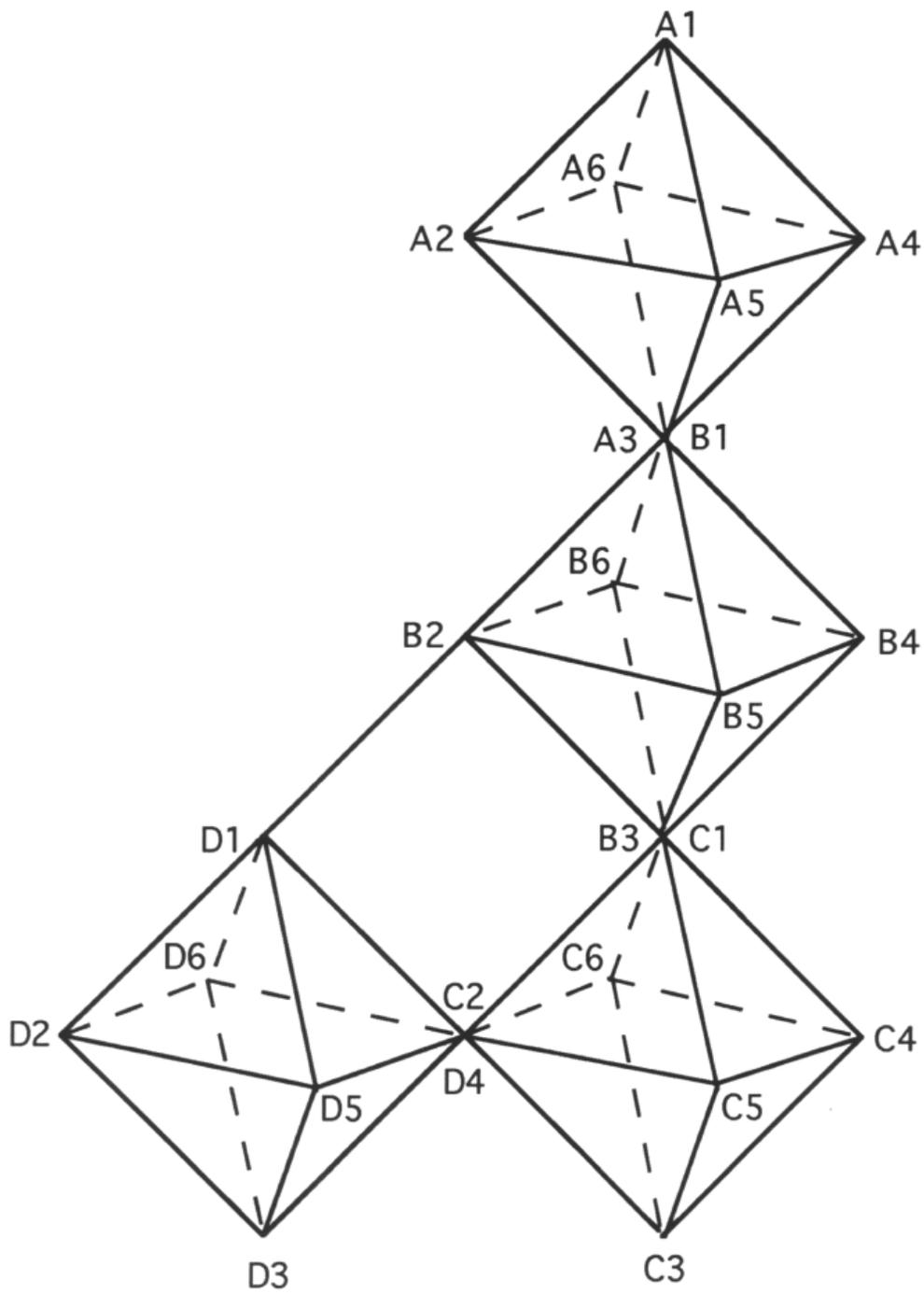
\end{document}